\documentclass[pra,twocolumn]{revtex4-1}
\usepackage[colorlinks,bookmarks=false,citecolor=blue,linkcolor=red,urlcolor=black]{hyperref}
\usepackage{graphicx}
\usepackage{amssymb,amsmath} 
\usepackage{amsfonts}
\usepackage{color}
\usepackage[utf8]{inputenc}
\usepackage[T1]{fontenc}
\usepackage{epsfig}
 
 \def\urlprefix{}
 \def\url#1{}
\def\be{\begin{equation}}
\def\ee{\end{equation}}
\def\bea{\begin{eqnarray}}
\def\eea{\end{eqnarray}}
\def\bi{\begin{itemize}}
\def\ei{\end{itemize}}
\def\bin{\begin{enumerate}}
\def\ein{\end{enumerate}}

\begin{document}
\title{Role of correlations and off-diagonal terms in binary disordered one dimensional systems}
\author{Arkadiusz Kosior$^1$, Jan Major$^1$, Marcin P\l{}odzie\'n$^1$, and Jakub Zakrzewski$^{1,2}$} 
\affiliation{\mbox{$^1$Instytut Fizyki imienia Mariana Smoluchowskiego, Uniwersytet Jagiello\'nski, \L{}ojasiewicza 11, 30-348 Krak\'ow, Poland}
\mbox{$^2$Mark Kac Complex Systems Research Center, Jagiellonian University, \L{}ojasiewicza 11, 30-348  Krak\'ow, Poland}
}

\begin{abstract}
We investigate one dimensional tight binding model in the  presence of a correlated binary disorder. The disorder is due to the interaction of particles with  heavy immobile other species. Off-diagonal disorder is created by means of a fast periodic modulation of interspecies interaction. The method based on transfer matrix techniques allows us to calculate the energies of extended modes in the correlated binary disorder. We focus on $N$-mer correlations and regain known results for the case of purely diagonal disorder. For
off-diagonal disorder we find  resonant energies. We discuss ambiguous properties of those states and compare analytical results with numerical calculations. Separately we describe a special case of the dual random dimer model.
\end{abstract}
\maketitle
\section{Introduction}\label{sec:intro}

In one dimensional disordered systems particles cannot freely propagate. The particle undergoes a series of scattering processes and, as a result of the destructive interference, its density profile decays exponentially. This phenomenon is called the Anderson localization (AL) \cite{anderson58,Lagendijk2009}. AL  is a single particle effect and therefore cannot be observed directly in solids (because of the strong electron-electron and electron phonon interactions). 

Nevertheless, nowadays many different tight-binding models can be easily implemented and investigated with ultracold atoms in optical lattices. Ultracold atomic systems in optical lattices guarantee a high degree of controllability \cite{Bloch2008, Jaksch2005,Lewenstein12}, allowing to change experimentally parameters of the model, such as hopping amplitudes using periodic lattice modulations \cite{holthaus2005,Lignier2007,Struck2011}, or interparticle interaction strength using the magnetic Feshbach resonance \cite{Chin2010}. In particular, the interactions can be switched off completely, creating perfect conditions for the investigation of AL. Indeed, AL of matter waves has been already observed in the laboratory \cite{billy08,anderson3D} (as well as a similar Aubry-Andr\'{e} localization \cite{Aubry80,roati2008}). 

It is very convenient to study disorder phenomena with optical lattices. In particular, the diagonal disorder in ultracold systems can be very easily implemented by manipulating on-site energies, e.g. with incommensurate lattices \cite{roati2008} or interactions with other (heavy) species of particles \cite{gavish2003,Massignan06,Bongs_2006,Cirac2007,Graham2008}. The later method allows to introduce various types of correlations in the disorder distribution, e.g. dual random dimer models (DRDM), where two heavy atom never come in pairs) or $N$-mers, where heavy particles always occupy a series of neighboring $N$ sites. $N$-mer correlations have been studied in various systems such as photons in dielectric waveguide \cite{Peng2004,Zhao2007}, acoustic waves \cite{Barros2011} or electrons in crystalline lattice \cite{Evangelou1993}. The off-diagonal disorder arises quite naturally, as the differences in the on-site energies change the tunneling amplitude values. As such, the off-diagonal disorder is correlated 
with the diagonal disorder and its relative amplitude is small. On the contrary, in this paper we present a theoretical scheme allowing to control the off-diagonal disorder strength in an experiment (for the DRDM case see \cite{kosior15}).

In Section \ref{sec:model} we describe a tight binding model with the binary disorder in both on-site energies and tunneling strengths.  Also, the methods of creating such a system using ultracold atoms in optical lattices is sketched. In Section \ref{sec:nmer} we describe a method employing transfer matrices to find resonant energies in the systems with binary correlated disorder, and use it to calculate positions of all transparent states in the case of $N$-mers with no off-diagonal disorder. In  Section  \ref{sec:off} we extend our analysis to include the diagonal - off-diagonal correlated disorder. We derive an equation for the resonant energies, discuss the asymptotic behavior and compare the results with numerical calculations of the localization length. Furthermore, in Section \ref{sec:drdm} we describe a dual random dimer model (being a $1$-mer model with off-diagonal correlated disorder). Again, we find \cite{kosior15} transparent modes, present an approximate expression for the localization length 
and compare the results with the numerical calculations.
 In the Appendix, a detailed description of the Floquet formalism used to derive effective Hamiltonian with off-diagonal disorder is presented.

\section{Model \label{sec:model}}
Consider a non-interacting tight-binding  one dimensional Hamiltonian
\begin{align}\label{Hm}
  H_0= \sum_i\left(\epsilon_i n_i-( a^\dagger _i a_{i+1}+\mathrm{h.c.})\right),
\end{align}
where $a_i$($a^\dagger_i$) is an operator of particle annihilation (creation) at site $i$, $n_i=a^\dagger_i a_i$ a particle number operator and  $\epsilon_i$ are on-site energies (we set $\hbar=1$ and use tunneling amplitude as an energy scale). 
The binary disorder occurs when $\{\epsilon_i\}$ are random numbers and take only two values: $\epsilon_a$ or $\epsilon_b$. Because of its simplicity, the binary disordered models serve as a playground for studying the localization and transport properties.

To create the binary disorder in an optical lattice one can use two species of atoms repulsively interacting with each other. We allow only one fraction of atoms to move freely on the lattice, while the second one is trapped in the lattice sites (we call them \emph{frozen} and denote with a $f$ superscript). The dynamics of the system is described by a Hamiltonian
\be
  H= \sum_i\left(V_0 n^f_i  n_i-( a^\dagger _i a_{i+1}+\mathrm{h.c.})\right),
\ee
where we eliminated  interactions between the \emph{mobile} particles (experimentally, one can switch them off by the optical or microwave Feshbach resonance \cite{Chin2010} or use spin polarized fermions). As $n^f$ is fixed, we can treat $V_0 n^f_i$ as an effective on-site energy $\epsilon_i$. Furthermore, if the \emph{frozen} particles are fermions or strongly repelling (hard-core) bosons, then their occupation per a lattice site $n^f_i$ is either zero or one and hence the on-site energies take only two values $\epsilon_i\in\{0,V_0\}$. 
 
The strength of the interspecies interaction $V_0$ can be changed in an experiment by applying a proper magnetic field $B$, which  is close to the value of the Feshbach resonance. In our scheme, we choose the magnetic field to be time periodic $B(t) = B(t+T)$, in such a way that the interactions are changing harmonically:
\be
\epsilon_i \rightarrow \epsilon_i(t)=n^f_i(V_0+V_1\sin(\omega t)).
\ee
As changing the magnetic field in a vicinity of the Feshbach resonance leads to rapid changes of the interaction strength, the relative values of $V_0$ and $V_1$ can be chosen at will in a broad range of values. In this way we obtain a time-dependent Hamiltonian:
\be
H(t)=\sum_i \left( n^f_i(V_0+V_1\sin(\omega t)) n_i - (a^\dagger_i a_{i+1}+\mathrm{h.c.}) \right).
\ee

If the  modulation frequency $\omega$ is bigger than other energy scales in system ($\omega \gg 1$) then, by the means of the Floquet theory, one can average out fast oscillating terms and find an effective Hamiltonian governing the long term dynamics of the system (see Appendix):
\be
H_{eff}=\sum_i\left(\epsilon_i n_i-t_i( a^\dagger _i a_{i+1}+\mathrm{h.c.})\right),
\label{eq:H3}
\ee
where $\epsilon_i= n^f_i V_0$ and $t_i=\mathcal{J}_0\left(\frac{V_1}{\omega}(n^f_{i+1}-n^f_i)\right)$ is a renormalized hopping. For a binary case $n^f_i\in\{0,1\}$:
 \begin{equation}\label{eq:valet}
 \epsilon_i=\left\{\begin{matrix}
 0, & \mbox{ if } n^f_i = 0\\ 
 V_0, & \mbox{ if } n^f_i =1
\end{matrix}\right.,\quad
 t_i=\left\{\begin{matrix}
 1, & \mbox{ if } n^f_i = n^f_{i+1}\\ 
 t', & \mbox{ if } n^f_i \neq n^f_{i+1}
\end{matrix}\right. ,
\end{equation}
where $t'=\mathcal{J}_0\left(V_1/\omega \right)$ varies from 1 to about -0.4 (to the minimum of the Bessel function, $\min_x\mathcal{J}_0(x)$).

Due to Anderson theory of localization  in one dimensional system all states should be localized, unless disorder have some correlations.
In the binary disordered systems correlations can take a form of $N$-mers where \emph{frozen} particles always come in series of length $N$, or dual random dimer model (DRDM) where no two \emph{frozen} particles can appear on the adjacent sites.
Such correlations can be created by using several lattices with different lattice constants (as described for the case of the DRDM in \cite{vignolo2010}).

\section{Delocalized modes for $N$-mers \label{sec:nmer}}
A transfer matrix method is a very effective tool for the numerical computation of the localization length \cite{delande2011}.  The formalism can be also used to find analytically the delocalized modes that are appearing in systems with the correlated binary disorder. A transfer matrix for a  system described by \eqref{eq:H3} has a form:
\be
\left(\begin{array}{c}\psi_{i+1}\\ \psi_{i}\end{array}\right)=\left(\begin{array}{c c}\frac{\epsilon_i-E}{t_i} & -\frac{t_{i-1}}{t_i}\\ 1&0\end{array}\right)\left(\begin{array}{c}\psi_{i}\\ \psi_{i-1}\end{array}\right)
\equiv T_i\left(\begin{array}{c}\psi_{i}\\ \psi_{i-1}\end{array}\right),
\ee
where $E$ is an energy of a state. Iterating the procedure:
\be
\left(\begin{array}{c}\psi_{j+1}\\ \psi_{j}\end{array}\right)= T_j\cdot\ldots\cdot T_i \left(\begin{array}{c}\psi_{i}\\ \psi_{i-1}\end{array}\right)
\equiv T_i^j \left(\begin{array}{c}\psi_{i}\\ \psi_{i-1}\end{array}\right).
\ee

In our analysis, we consider systems with binary disordered correlations in a form of $N$-mers where \emph{frozen} particles always come in series of length $N$. All such blocks have the same internal structure - it is the distribution of them which is random. Therefore, one can study transfer matrix of single block ranging from $i+1$ to $i+N$ (i.e. $T^{i+N}_i=T_{i+N}\cdot\ldots\cdot T_{i+1}$) to get some insight in the transport properties of the whole system. For the case of an $N$-mer with uniform tunnelings ($t'=1$), such a matrix can be written explicitly:
\be\label{eq:TMnmer}
T^{i+N}_{i+1} \equiv T^{N}(\varepsilon)= \left(\begin{array}{c c}w_{N}(\varepsilon) & -w_{N-1}(\varepsilon)\\ w_{N-1}(\varepsilon)&-w_{N-2}(\varepsilon)\end{array}\right),
\ee
where $\varepsilon=E-V_0$ and $w_n(\varepsilon)$ is a polynomial of the $n$-th order given by a recursive formula:
\be\label{eq:recc}
w_n(\varepsilon)= -\varepsilon \,w_{n-1}(\varepsilon) -w_{n-2}(\varepsilon),
\ee
where $w_{-1}(\varepsilon)=0$ and $w_{-2}(\varepsilon)=-1$. Solving the equation \eqref{eq:recc} we can get an explicit expression:
\be
w_{m}(\varepsilon)=\sum_{k=0}^{\lfloor m/2\rfloor} \left( \begin{matrix} m-k\\ k \end{matrix} \right)(-1)^{k+m}\,\varepsilon^{m-2k}.
\ee

In particular, a transfer matrix for a system without disorder (for an arbitrary number of sites) reads
\be
\label{eq:TMfree}
T^{M}(E)= \left(\begin{array}{c c}w_{M}(E) & -w_{M-1}(E)\\ w_{M-1}(E)&-w_{M-2}(E)\end{array}\right).
\ee
We can compare a free transfer matrix \eqref{eq:TMfree} with the $N$-mer transfer matrix \eqref{eq:TMnmer} 
\begin{align}\label{eq:nmercond}
T^N(E-V_0)=T^M(E).
\end{align}
If the condition \eqref{eq:nmercond} is met for some energy $E_r$, then we can say that for a state with such an energy the $N$-mer disordered system has the same transport properties as a free system. Hence, this state is delocalized.
Such a procedure can be easily extended to any kind of correlations in a form of randomly distributed blocks with fixed internal structure, not necessarily $N$-mers.

A problem of finding transparent modes for binary disordered system with $N$-mer correlations can be fully solved by writing explicitly equations \eqref{eq:nmercond}, using \eqref{eq:recc} and the fact that determinant of  $T^N(\varepsilon)$ is one ($\det(T^N(\varepsilon))=1$). Then, it can be shown that the only nontrivial solutions (i.e. $V_0\ne0$) of the equation \eqref{eq:nmercond} exist if 
\be\label{eq:w_cond}
w_{N-1}(E-V_0)=0,
\ee
therefore a problem of finding resonant energies is reduced to finding zeros of the polynomial $w_{N-1}$. It immediately follows that:
\be
T^N(E_r-V_0)=\pm\left(\begin{array}{cc}1&0\\0&1\end{array}\right),
\ee
which means that \eqref{eq:nmercond} is trivially satisfied for $M=0$.
It turns our that a $N$-mer problem has $N-1$ resonant energies (this result, from a different approach, has been first presented in \cite{Izrailev1995b}):
\be\label{eq:Er}
E_r=V_0+2\cos(\frac{\pi}{N}i)\quad\mathrm{for}\; i\in\{1, \ldots,  N-1\}.
\ee
The additional condition for the appearance of extended states is that its energy has to lie in the energy band for a system without disorder ($|E_r|\leq 2$).

\section{Off-diagonal correlated disorder}\label{sec:off}
The properties of the system change dramatically if we add the off-diagonal disorder, as in \eqref{eq:H3}. The off-diagonal disorder arises naturally in the model, as the differences in the on-site energies alter the hooping terms, but usually their values are small comparing to the diagonal disorder. Nevertheless, in our approach we can independently manipulate the values of diagonal and off-diagonal disorder. For the $N$-mer, due to \eqref{eq:valet}, only the edge tunneling amplitudes are changed: $t_i=t_{i+N}=t'$ (we assume here that $N$-mers are separated by at least one empty site). To obtain a transfer matrix of the $N$-mer, we need to write a transfer matrix for $N+2$ sites (since the altered tunneling amplitudes $t'$ appear in $T_i$, $T_{i+1}$, $T_{i+N}$ and $T_{i+N+1}$ we have to include two extra matrices):
\be
\tilde{T}^N(\varepsilon)\equiv T_i^{i+N+1} =T_{i+N+1}T_{i+N} \underbrace{T_{i+N-1} \ldots T_{i+2}}_{T^{N-2}(\varepsilon)}T_{i+1}T_i,
\ee
resulting in (where we denote the matrix elements with $v_M$, $v_{M-1}$, $v_{M-2}$ for the later convenience)
\be
\tilde{T}^N(\varepsilon) =
\left[
\begin{matrix}
 v_M & -v_{M-1} \\
 v_{M-1} & -v_{M-2}
\end{matrix}
\right],
\ee
where
\begin{eqnarray} \label{eq:vn}
 v_{M-2} &=&\frac{1}{t'^2}w_N(\varepsilon) ,\\
 v_{N-1} &=& -\frac{E}{t'^2}w_{N}(\varepsilon)- w_{N-1}(\varepsilon), \nonumber  \\ 
 v_N &=& \frac{E^2}{t'^2} w_{N}(\varepsilon)+2E w_{N-1}(\varepsilon) + t'^2 w_{N-2}(\varepsilon). \nonumber 
 \end{eqnarray}
It is straightforward to check that a condition 
\be \label{eq:transfer_m_tylda}
\tilde{T}^N(\varepsilon)=T^M(E)
\ee
is not satisfied by energies given by \eqref{eq:Er}. To fully solve the equation \eqref{eq:transfer_m_tylda}, we use the same procedure as in Section \ref{sec:nmer}. We assume that \eqref{eq:vn} fulfill a recursive formula:
\be\label{eq:v_rec_eq}
v_n = -E v_{n-1} -v_{n-2}.
\ee
It is a necessary but not sufficient condition for the resonance, as we do not know if there exist $M\in\mathbb{Z}$ such that \eqref{eq:transfer_m_tylda} is satisfied (in other words, we do not know if initial conditions of \eqref{eq:v_rec_eq} meet that of \eqref{eq:recc}).

From \eqref{eq:v_rec_eq} we can get an equation for zeros of polynomial of $N$-th order in $\varepsilon$:
\be\label{eq:tunnelingsresonance}
((1-t'^2)\varepsilon-t'^2V_0)w_{N-1}(\varepsilon)+(1-t'^4)w_{N-2}(\varepsilon)=0.
\ee
First, we can analyze limiting cases: for $t'=1$ we restore the equation \eqref{eq:w_cond}, so the approach is consistent with the one presented in Section \ref{sec:nmer}. In the case of $t'\rightarrow\pm\infty$ we get $w_{N-2}(\varepsilon)=0$, so the resonances can appear for the same energies as for the $(N-1)$-mer with uniform tunnelings. The case of $t'=0$ is trivial, as the chain is broken into pieces, and hence no transport can occur. Still, we can find that \eqref{eq:tunnelingsresonance} comes down to $w_N(\varepsilon)=0$, so for $t'\rightarrow0$ the resonant energies approach those of the $(N+1)$-mer.

\begin{figure}
\includegraphics[width=8.5cm]{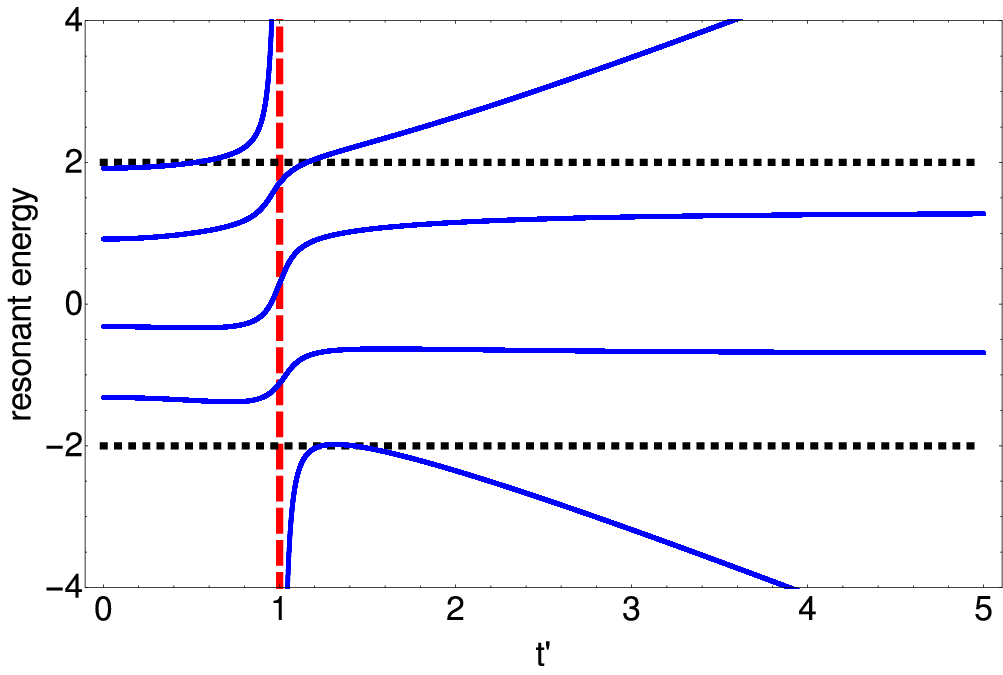}
\caption{(Color online) Blue solid lines: resonant energies for $4$-mer for $V_0=0.3$ (solutions of \eqref{eq:tunnelingsresonance}) in function of $t'$. Also band edges ($E=\pm2$) and $t'=1$ where solutions are given by \eqref{eq:Er} are indicated.}
\label{fig:er}
\end{figure}

\begin{figure}
\includegraphics[width=8.5cm]{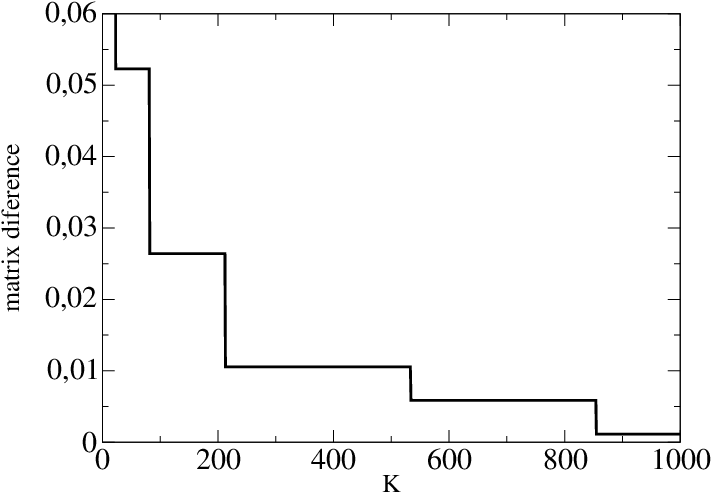}
\caption{(Color online) Asymptotic fall of matrix difference ($\min_{M\le K}\|\tilde{T}^N(\varepsilon)-T^M(E)\|_2$) to zero in function of $K$. Results for $4$-mer with $V_0$=0.5 and $t'=0.5$ for lowest resonant energy $E_R=-1.1485$.}
\label{fig:asympt}
\end{figure}

Solving \eqref{eq:tunnelingsresonance} numerically gives always $N$ solutions, for $t'\rightarrow1$ one of them diverges and the rest converges to $N-1$ solutions for the $N$-mer with uniform tunnelings. If $t'\rightarrow\infty$ two edge solutions diverge and the rest of them go asymptotically to the solutions for the $(N-1)$-mer. An example for the $4$-mer with $V_0=0.3$ is presented in Fig. \ref{fig:er}. 
Nevertheless, only for $t'=1$ we can say that states with $E_r$ move through the system in a way as there were no disorder. For $t'\neq1$ situation is ambiguous, the equation \eqref{eq:v_rec_eq} is satisfied, but we are unable to find $M$ for which $\tilde{T}^N(\varepsilon)=T^M(E)$. However, for $K\rightarrow\infty$, the difference asymptotically diminishes: 
\be
\min_{M\le K}\|\tilde{T}^N(\varepsilon)-T^M(E)\|_2\xrightarrow{K\rightarrow\infty} 0.
\ee
where $\|.\|_2$ is Frobenius matrix norm. Numerical results for the $4$-mer with $V_0=0.5$, $t'=0.5$ (for $E_R=-1.1485$) are shown in Fig. \ref{fig:asympt}.

\begin{figure}
\includegraphics[width=8.5cm]{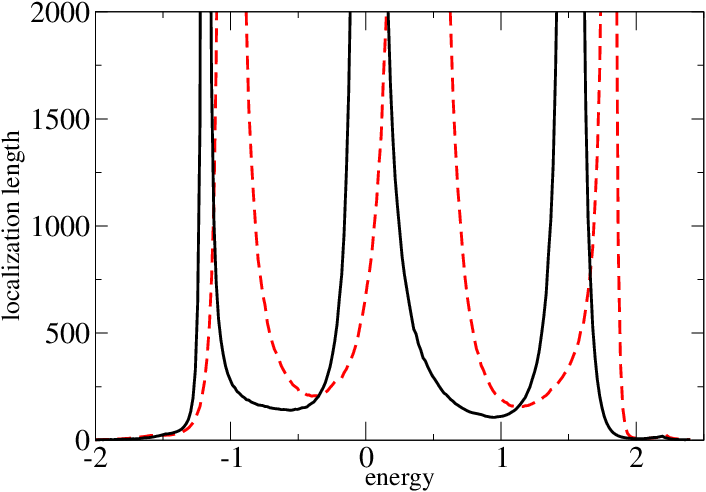}
\caption{(Color online) Anderson localization length obtained from transfer matrix calculations for 4-mer model with $V_0=0.4$ and mean occupation of frozen particles $0.55$. Red dashed line $t'=1$, black solid $t'=0.9$.}
\label{fig:plot1}
\end{figure}

In Fig.~\ref{fig:plot1} we present numerically calculated Anderson localization lengths (using the standard transfer matrix method, see e.g.  \cite{delande2011}) for the 4-mer and the mean occupation $0.55$, the strength of disorder is set to $V_0=0.4$. The red dashed line corresponds to the system with uniform tunnelings $t'=1$ while the black solid line is obtained for $t'=0.9$. Resonances in case of $t'=1$ are present at positions given by \eqref{eq:Er} while positions of resonances for $t'=0.9$ agree with those obtained using \eqref{eq:tunnelingsresonance}. Values of localization length in the vicinity of $E_r$ for the  case of $t'=0.9$ shows that the resonance indeed exists. Localization length seems to diverge as increasing the resolution makes it bigger.
\section{Dual random dimer model}\label{sec:drdm}
All the above analysis was valid for $N$-mers with $N\geq2$, the case of $1$-mer is special. For the case of uniform tunnelings $t'=1$ it does not have any delocalized modes --  which is obvious as it is just  the uncorrelated disorder. Even if we add an auxiliary condition that no two \emph{frozen} particles can occupy adjacent sites (which is equivalent to a condition of separation of $N$-mers), situation does not change -- it follows trivially from \eqref{eq:nmercond}. However if $t'\neq1$, we get the so called dual random dimer model (DRDM).  It has been extensively studied\cite{DRDM, vignolo2010,kosior15}. Such a model has the resonant energy
\be
E_r= \frac{V_0}{1-t'^2}, 
\label{rezonans_DRDM}
\ee
for which:
\be
\tilde{T}^1(E_r-V_0)=T^3(E_r)
\ee
is satisfied. As $M$ is finite,  this resonance is similar to resonances for $N$-mers with uniform tunnelings rather than for case with $t'\neq1$.
The resonance (\ref{rezonans_DRDM}) is present as long as the resonant energy lies withing the band width $|E_r|\le 2$
which is equivalent to the condition: $V_0 \le 2\left|1-t'^2 \right|$. 

The presence of the resonant mode results in the divergence of the AL length. In the limit of a small disorder (both diagonal $V_0\approx 0$ and off-diagonal $t'\approx 1$) it is possible to obtain an analytical expression for the inverse localization length \cite{kosior15}
\begin{eqnarray}   \label{loc_length}
 \lambda^{-1}& = & \frac{\rho}{(1+\rho)^2}\frac{(V_0+2 (1-t'^2)\cos (k))^2}{8\sin^2(k)}  \\
   & \times \nonumber & \left(1-2\frac{\rho(\rho+\cos(2k))}{1+\rho^2+2\rho\cos(2k)}\right),
\end{eqnarray}
where we substituted $E=2\cos(k)$ for an eigenenergy $E$, and  $\rho=\tilde{\rho}/(1-\tilde{\rho})$ where $\tilde{\rho}$ is the mean occupation number of \emph{frozen} particles. Indeed, the expression (\ref{loc_length}) vanishes for the resonant energy $E_r$.
\begin{figure}
\includegraphics[width=8.5cm]{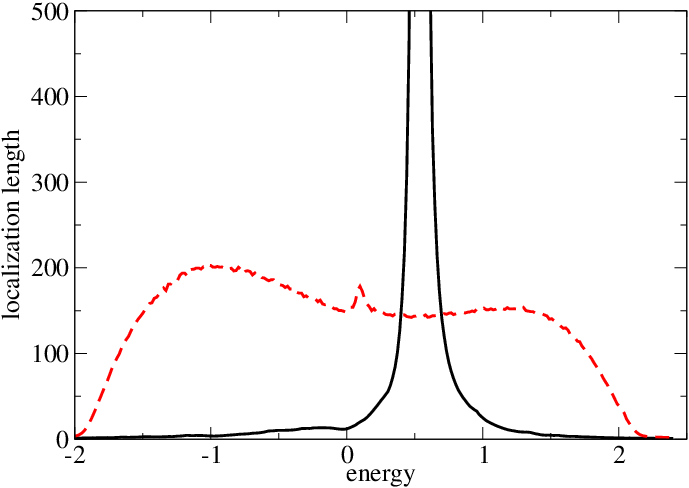}
\caption{(Color online) Anderson localization length obtained from transfer matrix calculations for 1-mer model with $V_0=0.4$ and mean occupation of frozen particles $0.55$. Red dashed line $t'=1$, black solid $t'=0.5$ (DRDM).}
\label{fig:plot2}
\end{figure}
\section{Conclusions}
In this work we have described a method of preparing the binary disorder in optical lattices using two atomic species. In such systems, the off-diagonal disorder can be created using the fast periodic modulation of the interspecies interactions. Consequently, effective values of diagonal and off-diagonal disorder can be changed independently in a broad range. In order to observe single particle localization, we assume that either our mobile particles are spin polarized fermions or that we switch off interactions using the optical Feshbach resonance.

We made a detailed analysis of the localization properties in systems with the disorder with correlations in a form of $N$-mers, especially determining the resonant energies for which the unobstructed transport through the system occurs. For that purpose, we employed the transfer matrix formalism. In this way, we restored a known result in the case of no off-diagonal disorder \cite{Izrailev1995b}. Furthermore, we extended the analysis to the case with disorder in tunnelings and derived an equation for the values of the resonant energies and analyze their asymptotic values.

We showed that, except the case of $t'=1$, initial conditions of recurrence equation \eqref{eq:recc} are met only asymptotically. It is not clear whether these states are extended,  but it seems that particles with the nearly resonant energies always escape finite systems.


\section*{Acknowledgments}
We acknowledge a support of the Polish National Science Centre via project DEC-2012/04/A/ST2/00088. AK acknowledges support in a form of a special scholarship of Marian Smoluchowski Scientific Consortium Matter Energy Future from KNOW funding. Support from the EU Horizon 2020-FET QUIC 641122 is also acknowledged.

\bibliographystyle{apsrev}

\appendix

\section{Fast modulation of on-site energies \label{app:1}}

In this section we derive an effective Hamiltonian for a cold atom realization of binary disorder with off-diagonal terms. The scheme we present allows to manipulate experimentally relative values of diagonal and off-diagonal disorder. 

In Section \ref{sec:model} we consider the on-site energy variations due to disorder caused by a \emph{frozen} fraction of particles. We apply periodic modulations of the on-site energies and obtain time dependent tight-binding Hamiltonian
\be\label{eq:Hamiltonian_time}
H(t)=\sum_i \left( n^f_i(V_0+V_1\sin(\omega t)) n_i - (a^\dagger_i a_{i+1}+\mathrm{h.c.}) \right).
\ee
As long as the driving frequency $\omega$ is large ($\omega\gg1$) then we can expect that there exists an effective time-independent Hamiltonian $H_{eff}$ which governs the long term dynamics. The explicit form of $H_{eff}$ can be found very rigorously using the well established formalism of Floquet theory \cite{floquet1883,Shirley1965,holthaus2005,Lignier2007,Struck2011,Rahav2003,eckardt15,bukov15}.

The Floquet theorem, being a time analogue of the Bloch theorem, works for time periodic Hamiltonians $H(t)=H(t+T)$. In this case, the time dependent Schr\"{o}dinger equation
\be
i\partial_t|\psi_n(t)\rangle=H(t)|\psi_n(t)\rangle
\ee
has solutions of a form
\be
|\psi_n(t)\rangle=e^{-i\varepsilon_n t}|u_n(t)\rangle
\ee
known as the Floquet states. Functions $|u_n(t)\rangle$ are $T-$periodic and $\varepsilon_n$ is called the quasienergy. Although $|\psi(t)\rangle$ are not eigenstates of $H(t)$, it appears that $|u(t)\rangle$ are  eigenstates of the Floquet Hamiltonian 
\be\label{eq:floquet_ham}
\mathcal{H}(t) = H(t)-i\partial_t
\ee
existing in the extended space of $T$-periodic functions. States in the extended space can be numbered using a new quantum number $m\in\mathbb{Z}$: $|u^m_n\rangle=|u_n\rangle e^{i\omega m t}$, where $|u_n^m\rangle$ is the eigenstate to the eigenenergy $\varepsilon^m_n=\varepsilon_n+\omega m$. It is straightforward to check that 
\be
e^{-i\varepsilon_n t}|u_n(t)\rangle=e^{-i\varepsilon_n^m t}|u_n^m\rangle=e^{-i\varepsilon_n^{m'} t}|u_n^{m
'}\rangle,
\ee
so $|\psi_n(t)\rangle$ does not depend on the choice of $m$ - adding $\omega$ to $\varepsilon_n$ in physical space is equivalent to going into the next "Brillouin zone"  for  the quasienergies. Therefore, an eigenvalue problem for Hamiltonian \eqref{eq:floquet_ham} can be reduced to a single "Brillouin zone", as long as we find its block-diagonal form in $m$-ordered basis.

Let us perform a unitary  transformation $U$:
\begin{align}\label{unitar}
\mathcal{H}'=U\mathcal{H}U^\dagger, \quad U=\exp{\left[-i \delta \frac{\cos(\omega t)}{ \omega}\sum_i \epsilon_i n_i\right]},
\end{align}
\be
\mathcal{H}'(t) =  \sum_i\left(\epsilon_i n_i-(e^{i(\epsilon_{i+1}-\epsilon_i)\frac{\delta \cos(\omega t)}{\omega}}a^\dagger _i a_{i+1}+\mathrm{h.c.})\right).
\ee
Although after the transformation $U$ a Hamiltonian matrix 
\be
\left[\mathcal{H}'\right]_{n',m';n,m}=\langle u^{m'}_{n'} |\mathcal{H}'|u^m_n\rangle
\ee
is not block diagonal, it turns out that coupling between different Floquet blocks ($m\ne m'$) can be neglected in the high frequency limit ($\omega \gg 1$). Consequently, we can consider only one diagonal block which is responsible for the long term ($t\gg 1/\omega$) dynamics:
\begin{align}\label{eq:H:PM1}
 H_{\mathrm{eff}}=\frac{1}{T}\int_0^{\frac{2\pi}{\omega}}dt\mathcal{H}'(t)=\sum_i\left(\epsilon_i n_i - (t_i a^\dagger _i a_{i+1}+\mathrm{h.c.})\right), 
\end{align}
where 
\begin{align} \label{tunel}
t_i=\mathcal{J}_0\left(\frac{\delta}{\omega}(\epsilon_{i+1}-\epsilon_{i})\right)
\end{align}
is the effective position dependent hopping and $\mathcal{J}_0$ is the zero-th order Bessel function. 

It is worth noting that $U$ adds only local phases to single particles states and does not alter the density distribution.  Therefore, localization properties of Hamiltonians \eqref{eq:Hamiltonian_time} and \eqref{eq:H:PM1} are the same. Moreover, if the change of parameters is sufficiently slow, then it is usually possible to adiabatically pass from eigenstates of one $H_\mathrm{eff}$ to another \cite{Poletti2011}. In particular, an eigenstate will follow the change of the disorder parameters during the experiment.

\end{document}